\documentstyle[12pt]{article}
\def\be{\begin{eqnarray}}
\def\ee{\end{eqnarray}}
\def\ba{\begin{array}}
\def\ea{\end{array}}
\def\p{\phi}
\def\vp{\varphi}
\def\a{\alpha}
\def\ep{\epsilon}
\def\pa{\partial}
\def\M{{\cal M}}
\def\G{{\cal G}}
\def\B{{\cal B}}
\def\A{{\cal A}}
\def\X{{\cal X}}
\def\H{{\cal H}}

\def\L{{\cal L}}
\def\E{{\cal E}}

\def\C{{\cal C}}
\def\D{^{(D)}}
\begin{document}
%{\large
%\draft
%\title
\begin{center}
{\Large \bf {
Israel--Wilson--Perj\'es Solutions \\
in Heterotic String Theory}}
\end{center}
%\author{}
\vskip 1cm
%\author
\begin{center}
{\bf \large \it {Alfredo Herrera-Aguilar}}
\end{center}
%\address{
\begin{center}
Joint Institute for Nuclear Research,\\
Dubna, Moscow Region 141980, RUSSIA.\\
e-mail: alfa@cv.jinr.dubna.su
\end{center}
%}
\vskip 0.5cm
%\author
%{
\begin{center}
and
\end{center}
%}
\vskip 0.5cm
%\author
\begin{center}
{\bf \large \it {Oleg Kechkin}}
\end{center}
%\address{
\begin{center}
Institute of Nuclear Physics,\\
M.V. Lomonosov Moscow State University, \\
Moscow 119899, RUSSIA, \\
e-mail: kechkin@monet.npi.msu.su
\end{center}
%}

%\maketitle

%%%%%%%%%%%%%%%%%%%%%%%%%%%%%%%%%%%%%%%%%%%%%%%%%%%%%%%%%%%%%%%%%%%%%%%%%%%%%
%\draft
\begin{abstract}
We present a simple algorithm to obtain solutions that generalize the
Israel--Wilson--Perj\'es class for the low-energy limit of heterotic string
theory toroidally compactified from $D=d+3$ to three dimensions. A remarkable
map existing between the Einstein--Maxwell (EM) theory and the theory
under consideration allows us to solve directly the equations of motion
making use of the matrix Ernst potentials  connected
with the coset matrix of heterotic string theory \cite {hk1}.
For the particular case $d=1$ (if we put $n=6$, the resulting
theory can be considered as the bosonic part of the action of $D=4$, $N=4$
supergravity) we obtain explicitly a dyonic solution
in terms of one  real $2\times 2$--matrix harmonic function and $2n$ real
constants ($n$ being the number of Abelian vector fields). By studying the
asymptotic behaviour of the field configurations we define the charges of the
system. They satisfy the Bogomol'nyi--Prasad--Sommmerfeld (BPS) bound.

\end{abstract}
%%%%%%%%%%%%%%%%%%%%%%%%%%%%%%%%%%%%%%%%%%%%%%%%%%%%%%%%%%%%%%%%%%%%%%%%%%%%%%
%\draft
%%%%%%%%%%%%%%%%%%%%%%%%%%%%%%%%%%%%%%%%%%%%%%%%%%%%%%%%%%%%%%%%%%%%%%%%%%%%%%
\newpage
%%%%%%%%%%%%%%%%%%%%%%%%%%%%%%%%%%%%%%%%%%%%%%%%%%%%%%%%%%%%%%%%%%%%%%%%%%%%
\section{Introduction}

Recently much work has been devoted to the construction of stationary
solutions for the low energy limit of heterotic string theory (or to
equivalent extended  supergravity models) \cite{kkot}--\cite{bko}.
These solutions were obtained by using duality
symmetries, by taking as ansatz harmonic functions
or by directly solving the associated Killing spinor
equations in the supersymmetric case.
In this paper we obtain a class of extremal stationary solutions
by directly solving the equations of motion of the theory. 
In order to characterize these classes of solutions can be
used two approaches: the criteria of unbroken supersymmetries \cite {bko}
or the criteria of saturation of the BPS bound \cite{gh}--\cite{g1}.
Classical
solutions that saturate this bound are stable and usually associated with
solitons. In the framework of heterotic string theory, the BPS bound was
studied in \cite{ss}--\cite{dlr}. The importance of BPS bounds in extended
supergravities was pointed out by Gibbons in \cite{g1}; it is related
to the existence of unbroken supersymmetry \cite{g2}. Since quantum
corrections to supersymmetric backgrounds are controllable, the BPS
solutions constitute an important tool to investigate the underlying
quantum theory from a non--perturbative point of view \cite{kall}.

The study of duality symmetries of compactified string theories can yield
non--perturbative information about the full string theory (for a review see 
for example \cite{fl} and references therein). These
symmetries relate different backgrounds which define essentially the same
quantum conformal field theories.

In this work we will make use of another kind of ``duality" between the
effective action of the low--energy limit of heterotic string theory
toroidally compactified to three dimensions and the stationary EM theory
(below we will present the map existing between both theories). Actually,
this is not a duality in the usual sense of the word, but the similarity
of the group structure of both theories enables us to express the effective
action of the heterotic string, as well as the equations of motion, in an 
EM form. We see that this fact also allows us to obtain exact solutions that 
generalize the solutions for the EM theory.
On the other hand, the effective action of heterotic string theory can be 
considered as a generalization of the EM one (if we set to zero the dilaton 
and axion fields `in the action', we get the action for the EM theory).
Thus, the study of the hidden symmetries that arise from dimensional 
reduction of EM theory provides many information about the rich
internal symmetries of heterotic string theory. A further reduction to
two dimensions leads to completely integrable theories (see for instance
\cite{m}--\cite{e...}), which entails infinite dimensional symmetries
(the Geroch group comes into play).

This letter is organized as follows: in Sec. 2 we summarize the
three--dimensional effective action of heterotic string compactified on a
seven torus. In Sec. 3 we express the action and the equations of motion of
the theory in terms of a pair of matrix Ernst potentials. In Sec. 4 we
indicate the procedure to obtain stationary extremal solutions
that generalize the IWP class for the effective action of the low--energy
limit of heterotic string theory. In Sec. 5 we obtain explicitly a BPS
saturated dyonic solution for the simplest case of the formalism: $d=1$.
There we define as well the set of charges of the system and show that they
satisfy the BPS bound. In Sec. 6 we summarize our results and discuss on
their implications.

%%%%%%%%%%%%%%%%%%%%%%%%%%%%%%%%%%%%%%%%%%%%%%%%%%%%%%%%%%%%%%%%%%%%%%%%%%%%
\section{Compactification to Three Dimensions}
Our starting point is the effective field theory of heterotic string in $D$
dimensions. The action of this theory describes gravity coupled to matter
fields \cite{ms}--\cite{as3}:
\be
S\D = \int d\D x \mid G\D\mid^{\frac{1}{2}}\,e^{-\p\D}(R\D &+& \p\D_{;M}
\p^{(D);M} - \frac{1}{12}\,H\D_{MNP}\, H^{(D)MNP}
\nonumber
\\
&-& \frac{1}{4}\,F^{(D)I}_{MN}\, F^{(D)IMN}),
\ee
where
\be
&&F^{(D)I}_{MN}=\pa _MA^{(D)I}_N-\pa _NA^{(D)I}_M,
\nonumber
\\
&&H\D_{MNP}=\pa _MB\D_{NP}-\frac{1}{2}A^{(D)I}_M\,F^{(D)I}_{NP}+
\mbox{\rm cycl. perms. of M,N,P.}
\nonumber
\ee
Here $G\D_{MN}$ is the $D$-dimensional metric, $B\D_{MN}$ is the
anti--symmetric Kalb-Ramond field, $\p\D$ is the dilaton and $A^{(D)I}_M$
denotes a set ($I=1,\,2,\,...,n$) of $U(1)$ gauge fields. At a generic
point of the moduli space, vector fields should form an Abelian
multiplet of dimension $d+n$, where $n$ is the number of initial vector
fields and $d$ is the number of compactified dimensions (in order to get a
self--consistent theory we must put $D=10$ and $n=16$ \cite {as3},
although we shall leave these parameters arbitrary for the sake
of generality).

In \cite {ms}-\cite {as3} it was shown that after the Kaluza-Klein
compactification of $d=D-3$ dimensions on a torus, the resulting theory is
\be
S^{(3)} = \int d^3 x \mid g\mid^{\frac{1}{2}}\,[R + \p_{;\mu} \p^{;\mu} &-&
\frac{1}{12}\,H_{\mu\nu\rho}\, H^{\mu\nu\rho}
-e^{-2\p}F^T_{\mu\nu}M^{-1}F^{\mu\nu} \nonumber
\\&-&\frac{1}{8}\,{\rm Tr}\left(J^M\right)^2].
\ee
Here the symmetric matrix $M$ has the following structure
\be
M=\left(
\ba{ccc}
G^{-1} & G^{-1}(B+C) & G^{-1}A \cr
(-B+C)G^{-1} & (G-B+C)G^{-1}(G+B+C) & (G-B+C)G^{-1}A \cr
A^{T}G^{-1} & A^{T}G^{-1}(G+B+C) & I_n+A^{T}G^{-1}A
\ea
\right)
\ee
with block elements defined by
\be
G=(G_{pq} \equiv G\D_{p+2,q+2}), \qquad
B=(B_{pq} \equiv B\D_{p+2,q+2}), \qquad
A=(A^I_p \equiv A^{(D)I}_{p+2}),
\ee
$C=\frac{1}{2}AA^{T}$ and $p,q=1,2,...,d$. Matrix $M$ satisfies the
$O(d,d+n)$ group relation
\be
M^TLM=L, \quad
{\rm where} \quad
L=\left(
\ba{ccc}
O & I_d & 0  \cr
I_d & 0 & 0  \cr
0 & 0 & -I_n
\ea
\right);
\ee
thus $M\in O(d,d+n)/O(d)\times O(d+n)$.

The remaining $3$-fields are defined in the following way:
for dilaton and metric fields one has
\be
\p=\p\D-\frac{1}{2}{\rm ln\,|det}\,G|, \qquad
g_{\mu\nu}=e^{-2\p}\left(G\D_{\mu\nu}-G\D_{p+2,\mu} G\D_{q+2,\nu}G^{pq}
\right).
\nonumber
\ee
Then, the set of Maxwell strengths $F^{(a)}_{\mu\nu}$ ($a=1,2,...,2d+n$) is
constructed on the basis of $A^{(a)}_{\mu}$, where
\be
&&A^p_{\mu}=\frac{1}{2}G^{pq}G\D_{q+2,\mu} \quad
A^{I+2d}_{\mu}=-\frac{1}{2}A^{(D)I}_{\mu}+A^I_qA^q_{\mu}, \nonumber \\
&&A^{p+d}_{\mu}=\frac{1}{2}B\D_{p+2,\mu}-B_{pq}A^q_{\mu}+
\frac{1}{2}A^I_{p}A^{I+2d}_{\mu}.
\nonumber
\ee
Finally, the $3$--dimensional axion
$$
H_{\mu\nu\rho}=\pa_{\mu}B_{\nu\rho}+2A^a_{\mu}L_{ab}F^b_{\nu\rho}+
\mbox{\rm cycl. perms. of $\mu$, $\nu$, $\rho$}
$$
depends on the $3$-dimensional Kalb-Ramond field
$$
B_{\mu\nu}=B\D_{\mu\nu}-4B_{pq}A^p_{\mu}A^q_{\nu}-
2\left(A^p_{\mu}A^{p+d}_{\nu}-A^p_{\nu}A^{p+d}_{\mu}\right).
$$

In three dimensions this system can be simplified because the Kalb-Ramond
field $B_{\mu\nu}$ has no physical degrees of freedom.
Moreover, the fields $A_\mu^a$ are dualized on-shell as follows
\be
e^{-2\p}MLF_{\mu\nu}=\frac{1}{2}E_{\mu\nu\rho}\nabla^{\rho}\psi;
\ee
so, the final system is defined by the quantities $M$, $\p$ and $\psi$. As
it had been established by Sen in \cite {as3}, it is possible to introduce the
matrix $\M_S$ in terms of which the action of the system adopts the
standard chiral form
\be
\ba{l}
S^{(3)}=
\int d^3 x \mid g\mid^{\frac{1}{2}}\,
\left[R - \frac{1}{8}\,Tr\left(J^{\M_S}\right)^2\right],
\ea
\ee
where $J^{\M_S}=\nabla\M_S\M^{-1}_S$.
This matrix is symmetric $\M_S=\M^T_S$ and satisfies the
$O(d+1,d+n+1)$-group relation
\be
\M_S\L_S\M_S=\L_S \quad
{\rm with} \quad
\L_S=\left(
\ba{ccc}
L & 0 & 0  \cr
0 & 0 & 1  \cr
0 & 1 & 0
\ea
\right),
\ee
so that $\M_S$ belongs to the coset $O(d+1,d+n+1)/O(d+1)\times O(d+n+1)$.

In \cite{hk1} it was constructed another chiral matrix
$\M \in O(d+1,d+n+1)/O(d+1)\times O(d+n+1)$ that possesses the same
structure that $M$ with block components $\G$, $\B$ and $\A$ of
dimensions $(d+1)\times (d+1)$, $(d+1)\times (d+1)$ and $(d+1)\times n$,
respectively. In order to obtain this matrix it was necessary to express
the column $\psi$ in the form
\be
L\psi=\left(
\ba{l}
u \cr
v \cr
s
\ea
\right),
\ee
where $u$ and $v$ are columns of dimension $d$, whereas the dimension
of the column $s$ is $n$. Then, the simple formulae
\be
\G=\left(
\ba{cc}
-e^{-2\p}+v^TGv & v^TG \cr
Gv & G
\ea
\right), \quad
\B=\left(
\ba{cc}
0 &  -w^T \cr
w & B
\ea
\right), \quad
\A=\left(
\ba{c}
s^T+v^TA \cr
A
\ea
\right),
\ee
where $w=u+Bv+\frac{1}{2}As$,
actually solved the task.  In terms of these three matrices the matter
part of the action reads
\be
S^{(3)}_{matter}=-\int d^3 x \mid g\mid^{\frac{1}{2}}\,
Tr\left\{\frac{1}{4}\left[\left(J^{\G}\right)^2-\left(J^{\B}\right)^2\right]+
\frac{1}{2}\nabla \A^T\,\G^{-1}\nabla \A\right\},
\ee
where $J^{\G}=\nabla \G \G^{-1}$ and $J^{\B}=\left(\nabla
\B+\frac{1}{2}(\A\nabla \A^T-\nabla \A\A^T)\right)\G^{-1}$, which exactly
corresponds to the chiral one
\be
S^{(3)}[\M]=-\frac{1}{8}\int d^3 x \mid g\mid^{\frac{1}{2}}\,
Tr\left(J^{\M}\right)^2
\ee
where the matrix $\M$ is defined by the block
components $\G$, $\B$ and $\A$ in the same way that the
matrix $M$ is defined by $G$, $B$ and $A$:
\be
\M=\left(
\ba{ccc}
\G^{-1} & \G^{-1}(\B+\C) & \G^{-1}\A \cr
(-\B+\C)\G^{-1} & (\G-\B+\C)\G^{-1}(\G+\B+\C) & (\G-\B+\C)\G^{-1}\A \cr
\A^{T}\G^{-1} & \A^{T}\G^{-1}(\G+\B+\C) & I_n+\A^{T}\G^{-1}\A
\ea
\right).
\ee
This matrix is symmetric  and satisfies the $O(d+1,d+n+1)$-group
relation
\be
\M\L\M=\L, \quad
{\rm where} \quad
\L=\left(
\ba{ccc}
0 & I_{d+1} & 0 \cr
I_{d+1} & 0 & 0 \cr
0 & 0 & -I_n \cr
\ea
\right),
\ee
so it belongs to the coset $O(d+1,d+n+1)/O(d+1)\times O(d+n+1)$.
%%%%%%%%%%%%%%%%%%%%%%%%%%%%%%%%%%%%%%%%%%%%%%%%%%%%%%%%%%%%%%%%%%%%%%%%%%%%%%
\section{Matrix Ernst Potentials}

At this stage one can introduce the matrix potential \cite{hk2}
\be
\X=\G+\B+\frac{1}{2}\A\A^T=\left(
\ba{cc}
-e^{-2\p}+v^TXv+v^TAs+\frac{1}{2}s^Ts&v^TX-u^T \cr
Xv+u+As&X
\ea
\right),
\ee
where the $d \times d$ matrix potential $X=G+B+\frac{1}{2}AA^T$
was defined for the first time by Maharana and Schwarz in the case
when $A=0$ \cite {ms}. Thus, the pair of potentials $\X$ and $\A$
allows us to express the $3$-dimensional action as follows
\be
S^{(3)} = \int d^3 x \mid g\mid^{\frac{1}{2}}\,\{-R &+& {\rm Tr}[
\frac{1}{4}\left(\nabla \X - \nabla \A\A^T\right)\G^{-1}
\left(\nabla \X^T-\A\nabla \A^T\right)\G^{-1} \nonumber \\
&+& \frac{1}{2}\nabla \A^T\,\G^{-1}\nabla \A]\},
\ee
where $\G=\frac{1}{2}\left(\X+\X^T-\A\A^T\right)$.
Its form is very similar to the stationary Einstein--Maxwell system
(see for instance \cite {m} and \cite {iw}). Actually the map
\be
\Phi\rightarrow i\A/\sqrt{2}, \qquad
\E\rightarrow \X
\ee
establishes a direct relation between EM theory and string
gravity, i.e., the matrix $\X$ formally plays the role of the
gravitational Ernst potential $\E$, whereas the matrix $\A$ corresponds
to the electromagnetic potential $\Phi$ of EM theory \cite{e}.
At the same time, one can notice a direct correspondence between the
transposition of $\X$ and $\A$ on the one hand, and the complex
conjugation of $\E$ and $\Phi$, on the other. This analogy
was useful to study the symmetry group of string gravity in \cite{hk1}.

The action (16) leads to the following equations of motion
\be
&&\nabla^2\X -2(\nabla \X-\nabla \A A^T)(\X+\X^T-\A\A^T)^{-1}\nabla \X=0,
\nonumber
\\
&&\nabla^2\A -2(\nabla \X-\nabla \A A^T)(\X+\X^T-\A\A^T)^{-1}\nabla \A=0.
\ee
It is worth noticing that these equations are quite symmetric each other
as in the case of EM theory.
%%%%%%%%%%%%%%%%%%%%%%%%%%%%%%%%%%%%%%%%%%%%%%%%%%%%%%%%%%%%%%%%%%%%%%%%%%%%%%
%%%%%%%%%%%%%%%%%%%%%%%%%%%%%%%%%%%%%%%%%%%%%%%%%%%%%%%%%%%%%%%%%%%%%%
\section{Generalized Israel--Wilson--Perj\'es Solutions}

In this Sec. we point out a simple algorithm to obtain real stationary
extremal (with a flat background metric) solutions for the equations of
motion (18) in terms of a real constant matrix of dimension $d+1$ and a
set of $2n$ real constants \cite{hk3}.
As in EM theory, we shall consider that there exists a linear dependence
between $\A$ and $\X$. Indeed, if one requires the matrix Ernst
potentials to satisfy the asymptotic flatness conditions
$\X_{\infty} \rightarrow \Sigma$, where
$\Sigma=diag(-1,-1,1,1,...,1)$, and $\A_{\infty} \rightarrow 0$,
the ``electromagnetic" potential can be expressed as follows
\be
\A=(\Sigma - \X)b\H,
\ee
where $b$ is an arbitrary constant $(d+1)\times n$--matrix and the
matrix $\H\in O(n)$ plays the role of a normalization factor which does
not affect the geometry of the theory since $\H\H^T=I_n$, we shall omit
this factor in what follows.
By performing the replacement (19) in the action (16) and setting the
Lagrangian of the system to zero (it implies that $R_{ij}=0$), we
get the following condition to be satisfied
\be
bb^T=-\Sigma /2;
\ee
it means that $b_{\tilde p n}b_{n\tilde q}=\delta_{\tilde p\tilde q}/2$
if $\tilde p,\tilde q=1,2$; and
$b_{\tilde p n}b_{n\tilde q}=-\delta_{\tilde p \tilde q}/2$ if
$\tilde p,\tilde q=3, 4,...,d+1$.  First of all
we can conclude that equation (20) is solvable if and only if
$n \geq d+1$, because for $n < d+1$, the equation system that
arises from it (for the components $b_{\tilde n}$) turns
out to be incompatible.

By substituting relation (19) in the equations of motion (18), {\it both}
of them reduce to the Laplace equation in Euclidean $3$--space
\be
\nabla ^2 [(\Sigma + \X)^{-1}]=0
\ee
which can be directly solved. For instance, we can immediately write
down a simple solution in terms of the harmonic function \footnote{This
solution correspond to a single object and can be simply generalized to a
multicenter solution by the following relation
$\frac{2}{\Sigma +\X}=\Sigma + \sum_j \frac{\tilde M_j}{r_j}$,
where $r_j=\sqrt{x_j^2+y_j^2+(z_j-i\a_j)^2}$ and the subscript $j$ labels
an arbitrary number of sources.}
\be
\frac{2}{\Sigma + \X}= \Sigma + \frac{\tilde M}{r},
\ee
where $\tilde M$ is a real $(d+1)$--dimensional arbitrary constant matrix,
$r=\sqrt{x^2+y^2+(z-i\a)^2}$ and $\a$ is a real constant. We choose $r$ in
this way in order to deal with rotating black hole solutions
\cite{bls}--\cite{sab} (in this case we have a ring singularity).
This is because general IWP solutions include both NUT charge and angular
momentum. We see that solution (22) is complex in general. On the other
hand, all variables entering the action (16) are real. In this paper
we shall restrict ourselves to a real class of solutions, leaving the study
of the complex solutions for the future. First of all we notice that we
could take into account the real or imaginary part of the harmonic function
(22); we shall put $1/R=Re 1/r$; secondly, we must require the constants
$\tilde M_{\tilde p\tilde q}$ and $b_{\tilde p n}$ to be real. However,
because of the indefinite character of $\Sigma$, matrix $b$ is complex in
general as well. We can procede as follows: Let us require just the first
two rows of $b$ to be real (leaving the remaining components imaginary),
then we perform the matrix product (19) and set the factors that multiply
the imaginary components of $b$ to zero. It turns out that this condition
imposes some restrictions on the matrix $\tilde M$ (some of its components
vanish) leading to real solutions for the potentials $\A$ and $\X$.

The procedure for obtaining the explicit form of the fields is the
following: once a solution of the Laplace equation is
written down (a solution of the form (22), for example), one solves the
system of matrix equations that arise from it and (19) in order to express
the $3$--fields $G$, $B$, $A$, $\phi$, $u$, $v$ and $s$ in terms of $R$
and the real arbitrary constants
$\tilde M_{\tilde p\tilde q}$ and $b_{\tilde p n}$.
Then one constructs the scalar column $\psi$ and
with the aid of the dualization
formula (6) one calculates the components of the ``true" gauge fields
$A^{(a)}_{\mu}$. This enables us to explicitly write down the $D$--fields
(see Sec. 2) as well as the $D$--dimensional interval.
%%%%%%%%%%%%%%%%%%%%%%%%%%%%%%%%%%%%%%%%%%%%%%%%%%%%%%%%%%%%%%%%%%%%%%%%%%%%%%%%%
\section{BPS Saturated Dyonic Solution}
In this Sec. we shall write down explicit solutions for the simplest case
of the formalism: $d=1$ (hence $n\geq 2$).
If we put $n=6$, the resulting theory can be considered as the bosonic part
of the action of $D=4$, $N=4$ supergravity \cite{bko}, where the axion field
can be introduced on--shell as follows:
\be
\pa _{\mu}a=\frac{1}{3}e^{-4\p}E_{\mu\nu\lambda\sigma}H^{\nu\lambda\sigma}.
\ee
where $E_{\mu\nu\lambda\sigma}=\sqrt{-g}\ep_{\mu\nu\lambda\sigma}$.
Thus, in this case $\tilde M$ becomes a $2\times 2$--matrix with $4$ real
parameters $m_{kl}$, $k,l=1,2$; and $b$ contains $2n$ real
constants constrained by the equality $b_{kn}b_{nk'}=\delta _{kk'}/2$.
So, we have at our disposal $2+2n$ (taking into account the parameter $\a$)
integration constants with $n\geq 2$ in the generic case.
On the other hand, the physical parameters of the theory are: The ADM mass,
the NUT, dilaton, and axion charges, the angular momentum and two sets of
$n$ electric and magnetic charges. There exist a relation between
the  charges of the theory to be satisfied: the Bogomol'nyi bound.
We see that the number of integration constants does not match the number of
physical parameters of the theory. This is because the condition
$b_{kn}b_{nk'}=\delta _{kk'}/2$ imposes $3$ constrains on the set of
electric and magnetic charges of the system. Thus, electric
and magnetic charges are not independent each other. Bellow we shall 
study the dependence that exists between them.

In \cite{bko} general SWIP solutions were obtained using as ansatz two
arbitrary harmonic functions. As we have said above, we suppose a linear
dependence between the matrix Ernst potentials. This fact allowed us to
reduce the equations of motion to a matrix Laplace equation and then to
write down the solution (22).

Since both theories are given in two very different settings, it is not easy
to make accurate comparisons of the obtained solutions. Another paper,
where we choose the solution of the Laplace equation in terms of two complex
harmonic functions, is in progress. In this case the settings of both
theories are more similar and we shall try to compare their solutions in a
simple way.

But let us keep in obtaining the solution, from (19) and (22) we have
\be
\X=
2Y-\Sigma= \left( \ba{cc} 2y_{11}+1 & 2y_{12} \cr
2y_{21} & 2y_{22}+1  \cr
\ea
\right)=
\left(
\ba{cc}
-e^{-2\p}+v^2X+vAs+\frac{1}{2}s^Ts&vX-u \cr
Xv+u+As&X
\ea
\right),
\nonumber
\ee
\be
\A=-2\left(Y+I_2\right)b=
\left(
\ba{c}
-2(y_{1k}+\delta_{1k})b_{kn} \cr
-2(y_{2k}+\delta_{2k})b_{kn} \cr
\ea
\right)=
\left(
\ba{c}
s^T+vA \cr
A
\ea
\right);
\ee
here we have introduced the matrix
$Y^{-1}=\left(\Sigma +\frac{\tilde M}{R}\right)$
with $det(Y^{-1})=1-\frac{m_{11}+m_{22}}{R}+\frac{det\tilde M}{R^2}\ne 0$.
We solve these equations and found the following expressions for the
$3$--fields
\be
G=-\frac{1-\frac{2m_{11}}{R}+\frac{m^2_{k1}}{R^2}}{\left(1-
\frac{m_{11}+m_{22}}{R}+\frac{det\tilde M}{R^2}\right)^2}=
-\frac{R^2(R^2-2m_{11}R+m^2_{k1})}{[R^2-(m_{11}+m_{22})R+det\tilde M]^2},\ \
e^{2\p}=1-\frac{2m_{11}}{R}+\frac{m^2_{k1}}{R^2},
\nonumber
\ee
\be
A=2\left(\frac{m_{21}R}{R^2-(m_{11}+m_{22})R+det\tilde M}b_{1n}+
\frac{m_{22}R-det\tilde M}{R^2-(m_{11}+m_{22})R+det\tilde M}b_{2n}\right),
\quad B=0,
\nonumber
\ee
\be
L\psi=
\left(
\ba{l}
u \cr
v \cr
s
\ea
\right)=
\left(
\ba{c}
(m_{12}-m_{21})R-m_{k1}m_{k2}/(R^2-2m_{11}R+m^2_{k1}) \cr
(m_{12}+m_{21})R-m_{k1}m_{k2}/(R^2-2m_{11}R+m^2_{k1}) \cr
2[(m_{11}R-m^2_{k1})b_{n1}+(m_{12}R-m_{k1}m_{k2})b_{n2}]/
(R^2-2m_{11}R+m^2_{k1}) \cr
\ea
\right).
\ee
After some algebraic calculations, from (6) we have
\be
\left(\nabla \times \overrightarrow A^{(a)}\right)^{\lambda}=
m^{(a)}\nabla^{\lambda}\left(\frac{1}{R}\right),
\ee
where $m^{(1)}=-(m_{12}-m_{21})/2$,\,\, $m^{(2)}=-(m_{12}+m_{21})/2$ \,\,
and \,\,$m^{(2+n)}=(m_{11}b_{n1}+m_{12}b_{n2})$.

The relation between physical parameters and integration constants
becomes evident when we switch from Cartesian to oblate spheroidal
coordinates defined by
\be
x=\sqrt{\rho^2+\a^2}sin\theta cos\vp,\quad
y=\sqrt{\rho^2+\a^2}sin\theta sin\vp,\quad
z=\rho cos\theta,
\ee
In terms of these coordinates the $3$--interval adopts the form
\be
ds^2_{3}=(\rho^2+\a^2cos^2\theta)(\rho^2+\a^2)^{-1}d\rho^2+
(\rho^2+\a^2cos^2\theta)d\theta^2+
(\rho^2+\a^2)sin^2\theta d\vp^2
\ee
and only the $A^{(a)}_{\vp}$ does not vanish\footnote{In fact we have
imposed the axial symmetry with respect to $z$.}:
\be
A^{(a)}_{\vp}=m^{(a)}cos\theta\frac{\rho^2+\a^2}
{\rho^2+\a^2cos^2\theta}=m^{(a)}\ep.
\ee
Studying their asymptotic behaviour we see that the integration constants
and the physical parameters of the theory are related by
\be
-g_{tt}=-G\sim 1+\frac{2m_{22}}{\rho}=1-\frac{2m}{\rho},
\nonumber
\ee
\be
\p\sim -\frac{m_{11}}{\rho}=\frac{D}{\rho},
\nonumber
\ee
\be
u\sim \frac{m_{12}-m_{21}}{\rho}=\frac{N_u}{\rho},
\nonumber
\ee
\be
v\sim \frac{m_{12}+m_{21}}{\rho}=\frac{Q_B}{\rho},
\nonumber
\ee
\be
s\sim 2\frac{m_{11}b_{n1}+m_{12}b_{n2}}{\rho}=
\frac{Q^{(n)}_m}{\rho},
\nonumber
\ee
\be
A^{(n)}_t\sim 2\frac{m_{21}b_{1n}+m_{22}b_{2n}}{\rho}=
\frac{Q^{(n)}_e}{\rho},
\ee
where the $m$ is the ADM mass of the black hole configuration, $D$, $N_u$,
$N_B$ are the dilaton, NUT and axion charges, respectively;
$Q^{(n)}_m$ and $Q^{(n)}_e$ are two sets of $n$ electric and magnetic
charges. These charges satisfy the BPS bound
\be
4(D^2+m^2)+2(Q^2_B+N^2_u)=\sum_n (Q_m^{(n)})^2+\sum_n (Q_e^{(n)})^2.
\ee

The complete solution is given by the following relations
\be
ds^2=G_{MN}dx^{M}dx^{N}=G\left(dt+\omega d\vp\right)^2+
e^{2\p}g_{\mu\nu}dx^{\mu}dx^{\nu},
\ee
where
\be
G=-\frac{(P^2-Q^2)\left[\rho^2-(m_{11}+
m_{22})\rho+m^2_{k1}-det\tilde M\right] (\rho^2-\a^2cos^2\theta)}
{(P^2+Q^2)^2}+
\nonumber
\ee
\be
\frac{2PQ\left[(3\rho^2-\a^2cos^2\theta)
(m_{22}-m_{11})+2\rho(m^2_{k1}-det\tilde M)\right]}{(P^2+Q^2)^2}-
\frac{P(\rho^2-\a^2cos^2\theta)+2\a cos\theta Q\rho}{P^2+Q^2},
\nonumber
\ee
\be
\omega =2A^{(1)}=2m^{(1)}\ep,
\nonumber
\ee
\be
e^{2\p}=\frac{(\rho^2+\a^2cos^2\theta)^2-2m_{11}\rho(\rho^2+\a^2cos^2\theta)+
m^2_{k1}(\rho^2-\a^2cos^2\theta)}{(\rho^2+\a^2cos^2\theta)^2},
\nonumber
\ee
\be
\p^{(4)}={\rm ln}\left(\frac{P(\rho^2-2m_{11}\rho+m^2_{k1}-\a^2cos^2\theta)+
2\a cos\theta Q (\rho-m_{11})}{P^2+Q^2}
\right),
\nonumber
\ee
\be
A^I_{\vp}=2\left(2\frac{(P\rho+Q)m_{21}b_{n1}+
(P(m_{22}\rho-det\tilde M)+m_{22}Q)b_{n2}}{P^2+Q^2} m^{(1)}-m^{(2+n)}
\right)\ep,
%\left(cos\theta+\a^2\frac{cos\theta sin^2\theta}
%{\rho^2+\a^2cos^2\theta}\right),
\nonumber
\ee
\be
A^I_{t}=2\frac{(P\rho+Q)m_{21}b_{1n}+
(P(m_{22}\rho-det\tilde M)+m_{22}Q)b_{2n}}
{P^2+Q^2},
\nonumber
\ee
\be
B_{t\vp}=2\left(m^{(2)}-\frac{(P\rho+Q)m_{21}b_{1n}+
(P(m_{22}\rho-det\tilde M)+m_{22}Q)_{b2n}}{P^2+Q^2} m^{(2+n)}
\right)\ep,
\nonumber
\ee
\be
B_{\mu\nu}=0,
\ee
where $P=\rho^2-(m_{11}+m_{22})\rho+det\tilde M-\a^2cos^2\theta$ and
$Q=(2\rho-(m_{11}+m_{22}))\a cos\theta$.

%%%%%%%%%%%%%%%%%%%%%%%%%%%%%%%%%%%%%%%%%%%%%%%%%%%%%%%%%%%%%%%%%%%%%%%%%%%%%%%%%
\section{Conclusions}
In this work we have presented a simple algorithm to obtain general exact
solutions that generalize the IWP class of EM theory for the effective action
of the low--energy limit of heterotic string
theory compactified to three dimensions in a $d=D-3$ torus. The method
allows one to reduce the motion equations to a matrix Laplace equation
just requiring the matrix Ernst potentials $\A$ and $\X$ to satisfy the
asymptotic flatness conditions (assuming a linear dependence
between them). Their solution is then expressed in terms of one
matrix harmonic function. For the simplest case of the formalism ($d=1$) a
BPS saturated dyonic
solution was explicitly constructed. This solution is parametrized by a
real $(2\times 2)$--matrix $\tilde M$, a set of $2n$ real constants $b_{kn}$
constrained by $b_{kn}b_{nk'}=\delta_{kk'}/2$ and the constant $\a$ which
defines the rotation of the configuration. The charges of the field system
saturate the Bogomol'nyi bound (31).

Among these solutions we identify rotating black hole--like solutions.
However, if we require the asymptotic flatness condition for the black hole
configuration to be satisfied, both the NUT parameter and the rotation one
vanish, leading to a static class of solutions.

%%%%%%%%%%%%%%%%%%%%%%%%%%%%%%%%%%%%%%%%%%%%%%%%%%%%%%%%%%%%%%%%%%%%%%%%%%%%%%
\section*{Acknowledgments}
We would like to thank our colleagues of DEPNI (NPI) and JINR for
encouraging us during the performance of this report. One of the authors
(A.H.) is grateful to N. Makhaldiani for helpful discussions and was
supported in part by CONACYT and SEP.
%%%%%%%%%%%%%%%%%%%%%%%%%%%%%%%%%%%%%%%%%%%%%%%%%%%%%%%%%%%%%%%%%%%%%%%%%%%%%%%
\newpage
%\begin{references}

%\end{references}

\end{document}